\documentclass[prl,twocolumn,amsmath,showpacs]{revtex4}
\input {epsf.sty}
\usepackage{epsfig}
\usepackage{amsmath,amsthm,amssymb}

\begin{document}
%\draft
\title{Collective Modes and $f$-wave Pairing
Interactions in Superfluid $^3$He}
\author{J.P. Davis, H. Choi, J. Pollanen, and W.P. Halperin}
\affiliation{Department of Physics and Astronomy,\\
       Northwestern University, Evanston, Illinois 60208}

\date{Version \today}


\begin{abstract}Precision measurements of
collective mode frequencies in superfluid $^{3}$He-B are sensitive
to quasiparticle  and $f$-wave pairing interactions. Measurements
were performed at various pressures using interference of
transverse sound in an acoustic cavity. We fit the measured
collective mode frequencies, which depend on the strength of
$f$-wave pairing and the Fermi liquid parameter $F_{2}^{s}$, to
theoretical predictions and discuss what implications these values
have for observing new order parameter collective modes.
\end{abstract}

\maketitle

\vspace{11pt} One of the fundamental notions of modern condensed
matter physics is that of spontaneous symmetry breaking at a phase
transition. A by-product of broken symmetry is the appearance of
order parameter collective modes, which are a fingerprint of the
underlying structure of the ordered state \cite{Sau81,Hal90}. We
can determine a wealth of information about the system through
measurement of these collective modes. Early observations of order
parameter collective modes in superfluid $^3$He provided
conclusive confirmation of its pairing state.  In addition to
better understanding the superfluid we can also learn about the
normal state quasi-particle interactions. The superfluid order
parameter has 9 complex valued components, resulting in a rich
spectrum of modes characteristic of the unconventional nature of
superfluidity in $^3$He. Similarly, one expects collective modes
in other unconventional pairing systems. In this work we report
precise measurements of the frequency of a collective mode in
superfluid $^{3}$He-B which depends, in part, on the strength of
sub-dominant $f$-wave pairing interactions.

In addition to the known $p$-wave pairing that defines the
equilibrium state of superfluid $^3$He there may be non-trivial
contributions from $f$-wave interactions \cite{Sau81} that appear
in the dynamics of the order parameter. Depending on the strength
of these interactions, Sauls and Serene \cite{Sau81} have
predicted the existence of modes with total angular momentum
$J=4$. The $f$-wave pairing is parameterized by,
\begin{equation}\label{x3}
    x_{3}^{-1} = \frac{1}{\ln\left(\frac{T_{cf}}{T_{c}}\right)},
\end{equation}
where $T_{c}$ is the transition temperature due to $p$-wave
interactions and $T_{cf}$ is the hypothetical transition
temperature due to $f$-wave interactions that would occur in the
absence of $p$-wave pairing.  This form for the $f$-wave
interaction parameter, $x_{3}^{-1}$, is chosen such that it is
zero if they are negligible, and large but negative if they are
significant and the $f$-wave pairing interaction is attractive.
Sauls and Serene have shown that this $f$-wave pairing influences
the frequency spectrum of collective modes, namely the real (+)
and imaginary (-) squashing modes labelled as $J=2^{\pm}$, each
having a total angular momentum classification of 2; the plus and
minus signs refer to the fact that different components of the
order parameter are involved in each case, either real or
imaginary. These squashing modes correspond to time-dependent,
anisotropic, momentum-space deformations (or squashings) of the
energy gap amplitude, $\delta\Delta^{+}(k,T)$.
\begin{figure}[b]
%%%%%%%%%%%%%%%%%   F I G U R E  1   %%%%%%%%%%%%%%%%%%
\centerline{\includegraphics[width=2.5in]{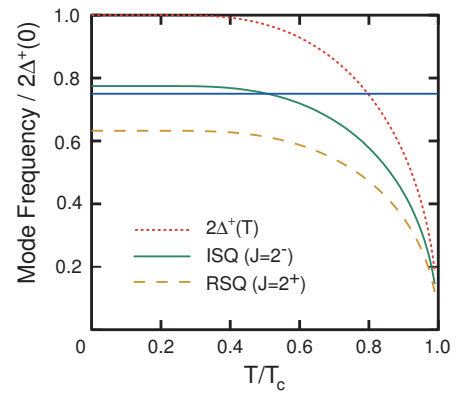}}
%%%%%%%%%%%%%%%%%%%%%%%%%%%%%%%%%%%%%%%%%%%%%
\caption {\label{fig1}(color online). Schematic of the collective
mode spectrum for $^3$He-\emph{B} normalized to twice the value of
the gap at zero temperature, $2\Delta^{+}(0)$.  From low to high
frequency we show the real squashing mode (dashed line), the
imaginary squashing mode (solid line), and $2\Delta^{+}(T)$
(dotted line), the threshold for pair-breaking. The straight line
is an example of a measurement frequency crossing the imaginary
squashing mode and pair breaking.}
\end{figure}
\noindent  In the absence of quasiparticle interactions and
$f$-wave pairing these modes have frequencies proportional to the
gap \cite{Hal90} which we take from the weak-coupling plus model
of Rainer and Serene \cite{Rai76}:
\begin{equation}\label{freq}
    \Omega_{2^{\pm}} = a_{\pm}\Delta^{+}(T,P),
\end{equation}
where $a_{+} = \sqrt{8/5}$ and $a_{-} = \sqrt{12/5}$. In general,
however, $a_{\pm}$ will have weak temperature and pressure
dependences. Fig.~\ref{fig1} is a schematic of these mode
frequencies as a function of temperature. To date, measurements of
the imaginary squashing mode (ISQ) frequency have not been as
precise as those of the real squashing mode (RSQ) and hence
determination of $x_3^{-1}$ from this mode has been inexact.  We
use a transverse acoustic cavity technique that takes advantage of
the existence of transverse acoustic standing waves near the
collective mode in order to measure presicely the ISQ mode
frequency.

The $^3$He sample is contained within a silver sample cell
attached to a nuclear demagnetization refrigerator \cite{Ham89}.
We used a lanthanum-diluted cerium magnesium nitrate paramagnetic
salt thermometer \cite{Ham89} calibrated with respect to the
superfluid phase diagram as given by Greywall \cite{Gre86}. Our
\emph{AC}-cut quartz transducers have coaxial electrode patterning
and an overtone polish.  An acoustic cavity, of spacing 48
microns, was formed from two transducers spaced by monodispersed
polystyrene latex microspheres. We obtained a frequency resolution
during experiments of $\Delta\nu / \nu \approx 2 \times 10^{-9}$
from 85 to 125 MHz (15$^{\mathrm {th}}$ to 21$^{\mathrm {st}}$
harmonics) using a CW, frequency-modulated, acoustic-impedance
spectrometer \cite{Ham89}.
\begin{figure}[t]
%%%%%%%%%%%%%%%%%   F I G U R E  2   %%%%%%%%%%%%%%%%%%
\centerline{\includegraphics[width=3.125in]{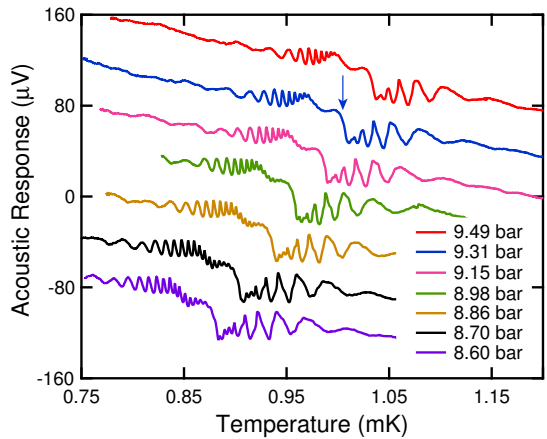}}
%%%%%%%%%%%%%%%%%%%%%%%%%%%%%%%%%%%%%%%%%%%%%
\caption {\label{fig2}(color online). Interference of transverse
sound near the imaginary squashing mode using 99.9484 MHz sound at
various pressures near 9 bar.  Note that as the pressure is
changed the frequency of the ISQ mode changes accordingly since
the gap amplitude, Eq.~\ref{freq}, is approximately proportional
to $T_{c}$. The cessation of wiggles and the bend in the acoustic
impedance marks the location of the imaginary squashing mode,
delineated by the arrow for 9.31 bar (see Fig.3).}
\end{figure}
\noindent

Transverse sound measurements using acoustic impedance techniques
in $^3$He are reviewed by Halperin and Varoquaux \cite{Hal90}. The
acoustic impedance is $Z=\rho/\omega q = \rho C$, where $C$ is the
complex phase velocity and $q = \omega / c + i\alpha$. Hence the
acoustic impedance is simultaneously sensitive to changes in the
(normal fluid) density, $\rho$; phase velocity, $c$; and
attenuation, $\alpha$.  Transverse sound has been observed as a
propagating mode in superfluid $^3$He \cite{Kal93,Lee99} following
the prediction of Moores and Sauls \cite{Moo93}, the only liquid
for which this has been demonstrated. As a result, the quartz
transducer, which forms one side of a cavity, detects an impedance
modulated by the  sound wave reflected from the opposite cavity
wall.  The interference between outgoing and reflected waves gives
rise to an oscillatory response in impedance as the velocity
changes near the mode frequency, Fig.~\ref{fig2}.

The acoustic cavity response as a function of temperature was
measured for a sequence of pressures at fixed acoustic frequency,
as shown for 99.9484 MHz near 9 bar in Fig.~\ref{fig2}. As the
temperature is lowered the ISQ frequency approaches, and then
crosses, the fixed transverse sound frequency (where the straight
line approaches the middle curve in Fig.~\ref{fig1}).
Correspondingly, according to the theoretical dispersion
represented in Eq.~\ref{dispersion}, the phase velocity increases
and transverse sound propagates with lower attenuation.  In fact,
it appears from this equation as though the velocity diverges at
the crossing if viewed sufficiently far from the crossing itself.
As the velocity increases so does the wavelength and for each
half-wavelength that leaves the cavity there is one oscillation.
As the ISQ mode frequency crosses the measurement frequency the
order parameter mode resonantly absorbs acoustic energy and the
interference pattern is extinguished. This corresponds to the
sharp upward bend with decreasing temperature in the traces in
Fig.~\ref{fig2}.  As the temperature is lowered further,
transverse sound propagates again and the interference pattern
reappears. Finally, as the temperature is lowered still further
the transverse sound becomes highly attenuated.  The details of
the oscillations on the low temperature side of the mode and the
acoustic impedance near the crossing point have not been observed
previously and are not predicted by Eq.~\ref{dispersion}.
\begin{figure}[b]
%%%%%%%%%%%%%%%%%   F I G U R E  3   %%%%%%%%%%%%%%%%%%
\centerline{\includegraphics[width=2.5in]{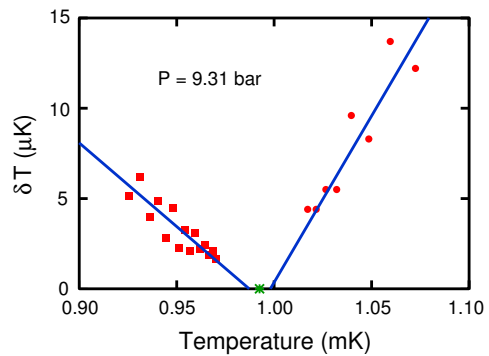}}
%%%%%%%%%%%%%%%%%%%%%%%%%%%%%%%%%%%%%%%%%%%%%
\caption {\label{fig3}(color online). Temperatures of maxima and
minima in impedance for determining the ISQ mode crossing at
99.9484 MHz.}
\end{figure}
\noindent

\begin{figure}[t]
%%%%%%%%%%%%%%%%%   F I G U R E  4   %%%%%%%%%%%%%%%%%%
\centerline{\includegraphics[width=3.125in]{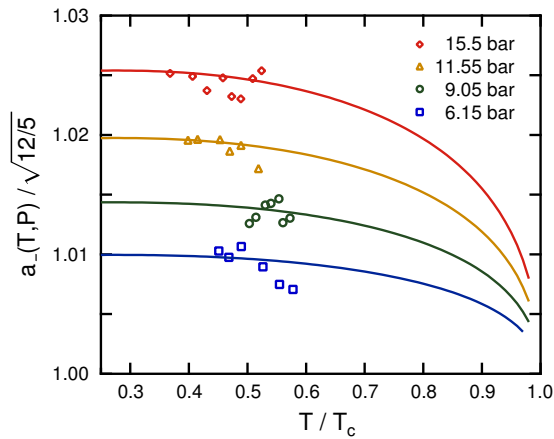}}
%%%%%%%%%%%%%%%%%%%%%%%%%%%%%%%%%%%%%%%%%%%%%
\caption {\label{fig4}(color online). Imaginary squashing mode
crossings as determined by the oscillation period shift analysis
of Fig.~\ref{fig3} corrected to be at constant pressures.  The
fits are to Eq.~\ref{interactions}.}
\end{figure}

In order to determine the exact crossing temperature for the
imaginary squashing mode with transverse sound, we plot, as in
Fig. 3 at 9.31 bar, the temperature difference of sequential
extrema in the interference pattern, $\delta T$, and extrapolate
independently from both the high and low temperature sides of the
ISQ.  The temperature at which $\delta T$ goes to zero, where the
velocity diverges, marks this crossing point.  The slopes of the
lines in the figure are determined by the temperature dependence
of the phase velocity of transverse sound. We find that they are
different on the high and low temperature sides by about a factor
of two. Our determinations of $a_{-}(T,P)/\sqrt{12/5}$ from the
mode crossings are plotted for four fixed pressures in
Fig.~\ref{fig4}. The crossings are determined with a resolution of
5 to 15 $\mu K$, resulting in an uncertainty in $a_{-}$ between
0.1\% and 0.25\%.
\noindent

These data can be compared with the theory for transverse sound
propagation in $^3$He as given by Moores and Sauls \cite{Moo93}.
They have shown that the dispersion of transverse sound is
\begin{multline}\label{dispersion}
    \left(\frac{\omega}{q v_f}\right)^{2} = \frac{F_1^{s}}{15}(1-\lambda) +
    \frac{2F_1^{s}}{75}\lambda\\
    \times\frac{\omega^{2}}{(\omega+i\Gamma)^{2} - \Omega_{2^{-}}^{2} -
\frac{2}{5}q^{2}v_f^{2}},
\end{multline}
where $v_{f}$ is the Fermi velocity, $\omega$ is the measurement
frequency, $\Omega_{2^{-}}= a_{-}(T,P)\Delta^{+}(T,P)$ is the ISQ
mode frequency, and $\lambda(\omega,T)$ is the Tsuneto function.
$\Gamma(T)$ is the width of the mode, with an approximate form of
$\Gamma(T)\simeq\Gamma_{c}\sqrt{T /
T_{c}}e^{-\frac{\Delta(T)}{T}}$ and $\Gamma_{c}\sim10^{6}-10^{7}$
Hz \cite{Moo93}. At temperatures low compared with $T_c$, the
first term goes to zero like the normal fluid density, and the
second term dominates. From Eq.~\ref{dispersion}, it can be seen
that transverse sound couples off-resonantly to the ISQ mode at
frequencies above $\Omega_{2^{-}}(T)$.  Below the ISQ mode, owing
to non-zero $\Gamma(T)$, transverse sound continues to propagate
but highly attenuated. The condition for the crossing represented
in Fig.~\ref{fig3} is shown by the vertical arrow in
Fig.~\ref{fig2}. This allows us to identify the position of the
ISQ mode as a specific point along the trace of acoustic
impedance. From Eq.~\ref{dispersion} we see that there are no
significant dispersion corrections to the mode frequency, a
consequence of the fact that transverse sound propagates only
through off-resonant coupling to the ISQ mode.

\begin{figure}[t]
%%%%%%%%%%%%%%%%%   F I G U R E  5   %%%%%%%%%%%%%%%%%%
\centerline{\includegraphics[width=3.125in]{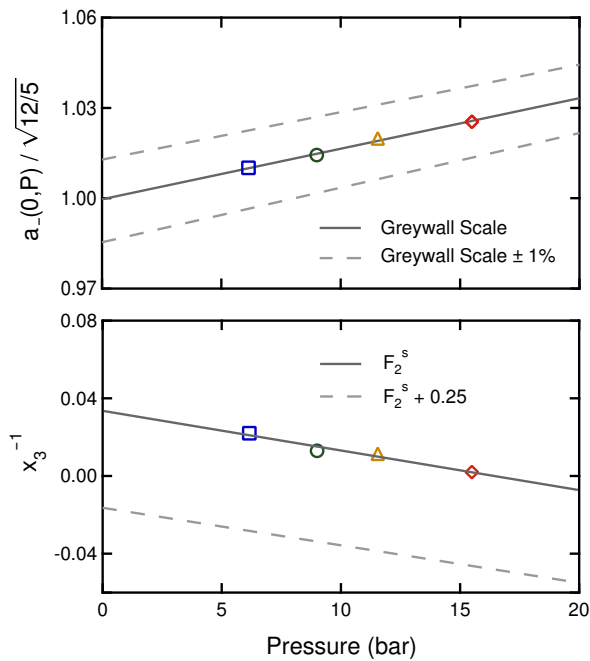}}
%%%%%%%%%%%%%%%%%%%%%%%%%%%%%%%%%%%%%%%%%%%%%
\caption {\label{fig5}(color online). Pressure dependence of (top)
$a_{-}(0,P)/\sqrt{12/5}$ and (bottom) $x_3^{-1}$ from the fits of
Eq.~\ref{interactions} to the ISQ mode frequencies. The straight
lines through the data  are guides to the eye. The open symbols
use the Greywall temperature scale \cite{Gre86}. The dashed lines
(top) correspond to a $\pm$1\% change to this scale.  The values
of $x_3^{-1}$  are obtained with $F_2^{s}$ from Ref. \cite{Hal90}.
The dashed line (bottom) shows the effect of adding 0.25 to
$F_2^{s}$.}
\end{figure}

The influence of quasiparticle interactions, $F_{2}^{s}$, and
$f$-wave pairing, $x_{3}^{-1}$, on the imaginary squashing mode
frequency, was calculated by Sauls and Serene \cite{Sau81},
\begin{multline}\label{interactions}
    \Omega_{2^{-}}^{2} - \frac{12}{5}\Delta^{+2} +
    \frac{3}{5}F_2^{s}(\Omega_{2^{-}}^{2} -
    4\Delta^{+2})\lambda\\
    +\frac{1}{4}x_{3}^{-1}\Omega_{2^{-}}^{2}(\Omega_{2^{-}}^{2}-4\Delta^{+2})\lambda/\Delta^{+2}= 0.
\end{multline}
If $F_2^{s}=0$ and $x_{3}^{-1}=0$ then Eq.~\ref{interactions}
reduces to Eq.~\ref{freq} and one would expect the data in
Fig.~\ref{fig4} to lie exactly at one, assuming correctness of the
temperature scale \cite{Gre86}. The curved lines in
Fig.~\ref{fig4} are fits to Eq.~\ref{interactions} using
$\lambda(\omega,T)$ adapted to the weak-coupling plus \cite{Rai76}
form of the gap, $\Delta^{+}(T,P)$. The only free parameter in
these fits is taken to be $x_{3}^{-1}$, with $F_2^s$ as an input
\cite{Hal90}. In the top of Fig.~\ref{fig5} we plot
$a_{-}(0,P)/\sqrt{12/5}$, the zero temperature intercepts of the
fits in Fig.~\ref{fig4}.  We find that at zero pressure
$a_{-}(0,0)/\sqrt{12/5}=1$. This is expected if quasiparticle
interactions become insignificant. There is reasonable evidence
that both $f$-wave pairing interactions and $F_2^{s}$ become small
at low pressure \cite{Hal90}.  However, any discrepancy between
the Greywall temperature scale \cite{Gre86} and the absolute
temperature scale will shift the data.  Possible inaccuracy in the
temperature scale was estimated by Greywall to be better than 1\%
\cite{Gre86}. Taking these extremes for the temperature scale as a
constant factor gives the dashed lines in the top panel of
Fig.~\ref{fig5}.  The level of spectroscopic precision is very
high. Nonetheless, we can at most state that the \emph{combined}
effect of quasiparticle interactions, $f$-wave pairing, and
inaccuracy in the temperature scale is negligible at $P=0$.

In the lower portion of Fig.~\ref{fig5} we show the pressure
dependence of the $f$-wave pairing strength, $x_{3}^{-1}$.  These
results hinge on the accuracy of $F_{2}^{s}$ derived from
measurements of the difference in first and zero sound velocities
tabulated in Ref. 2.  There is a reasonable consensus in published
data at intermediate pressures near 10 bar and above; although
less so at low pressure: \emph{i.e.} $F_{2}^{s}$ = 0.17, 0.34 and
0.5 at $P =$ 10, 15 and 20 bar and $F_{2}^{s} \approx 0$ at $P =
0$. Taking previous work \cite{Hal90} into account we estimate
that $F_{2}^{s}$ could be larger but likely not more than +0.25
for any pressure affecting our determination of $x_{3}^{-1}$ as
shown by the dashed line in Fig.~\ref{fig5}. In spite of these
inaccuracies it is nonetheless clear that the $f$-wave pairing
interaction becomes negative (attractive) with higher pressure.

\noindent

The values of $x_3^{-1}$ from our measurements can be compared
with those from other techniques.
 Acoustic RSQ measurements \cite{Fra89} yield values of $x_3^{-1}$
that roughly start at zero at zero pressure and increase to
$\sim-0.25$ at 20 bar.  These results depend on the Fermi liquid
parameter $F_2^{a}$ which is not well established. Analysis of the
acoustic Faraday effect \cite{Lee99} yields a value of $x_3^{-1}=
-0.375$ at 4.3 bar \cite{Sau99}. Meisel \emph{et al.} \cite{Mei87}
performed an analysis of longitudinal acoustic ISQ data to extract
$x_3^{-1}$. They concluded that $x_3^{-1}$ was $\sim0.2$ at 0 bar,
but decreased as the pressure increased becoming negative near 5
bar, then leveling off at $\sim-0.2$. The absolute values are
generally much larger than what we infer from our data but we note
that longitudinal sound measurements are inherently less precise
since this sound mode couples so strongly to the ISQ mode. We find
the pressure dependence of $x_{3}^{-1}$ to be about a factor of
six smaller than in these previous works independent of our
estimated inaccuracy in the temperature scale or in $F_{2}^{s}$.
We conclude that the predicted \cite{Sau81} $J=4$ modes may exist
but, if so, only very near or slightly below $2\Delta^{+}(T,P)$
\cite{Hal90}. Some of the modes in the 9-fold multiplet will
couple to transverse sound and our transverse acoustic cavity
technique should allow resolution of these collective modes at
high frequencies very close to the particle-hole continuum.  In
addition, application of a magnetic field will be helpful for
observation of $J=4$ modes whose frequencies decrease by the
Zeeman effect.

Fourier transform longitudinal acoustics experiments near the gap
edge were performed at low pressure by Masuhara \emph{et al.}
\cite{Mas00}. They observed the onset of anomalously high
attenuation at an energy (frequency) 4\% lower than was expected
from weak coupling BCS theory, $2\Delta_{BCS}$. Since it is
unlikely that the superfluid $^3$He-B order parameter is smaller
than $\Delta_{BCS}$, their results mean that either the
temperature scale is imprecisely known or there is a new
collective mode near the gap edge.  Our measurements of the ISQ
mode rule out that inaccuracy in temperature of this magnitude is
a possible explanation and suggest that these authors have
observed attenuation from higher order $J=4$ order parameter
collective modes.

In summary, we have made high precision measurements of an order
parameter collective mode using interference of transverse sound
in an acoustic cavity.  Taking into account strong coupling
effects we interpret our data to determine values of the $f$-wave
pairing interaction strength that are much smaller than those from
previous work.  Despite inaccuracy in the parameters required for
the analysis that reduce our resolution, we have set limits on the
pressure dependence of the strength of $f$-wave interactions in
superfluid $^3$He.  Increased accuracy in measurements of $F_2^s$
will help to refine this conclusion. Lastly, the high resolution
of the transverse acoustic cavity technique should make it
possible to observe the predicted $J=4$ order parameter collective
modes.

 We acknowledge support from the National
Science Foundation, DMR-0244099.  We would also like to thank J.A.
Sauls for helpful discussions.

\end{document}